\documentclass[conference]{IEEEtran}
\IEEEoverridecommandlockouts
\usepackage{cite}
\usepackage{amsmath,amssymb,amsfonts}
\usepackage{algorithmic}
\usepackage{graphicx}
\usepackage{textcomp}
\usepackage{xcolor}

\def\BibTeX{{\rm B\kern-.05em{\sc i\kern-.025em b}\kern-.08em
    T\kern-.1667em\lower.7ex\hbox{E}\kern-.125emX}}
\begin{document}

\title{Att-KGCN: Tourist Attractions Recommendation System by using Attention mechanism and Knowledge Graph Convolution Network\\
{\footnotesize \textsuperscript{}}
}
\author{ \IEEEauthorblockN{1\textsuperscript{st} Ahmad A. Mubarak}
\IEEEauthorblockA{\textit{College of Applied Sciences and Educational} \\
\textit{Ibb University}\\
Ibb, Yemen \\
ahmedmubarak@ibbuniv.edu.ye}
\and
\IEEEauthorblockN{2\textsuperscript{nd} JingJing Li }
\IEEEauthorblockA{\textit{School of Computer Science} \\
\textit{Shaanxi Normal University}\\
Xi'an, China \\
ljjhpu@163.com}
\and
\IEEEauthorblockN{3\textsuperscript{ed}  Han Cao}
\IEEEauthorblockA{\textit{School of Computer Science} \\
\textit{Shaanxi Normal University}\\
Xi'an, China \\
 caohan@snnu.edu.cn}

}
\maketitle

\begin{abstract}
The recommendation algorithm based on knowledge graphs is at a relatively mature stage. However, there are still some problems in the recommendation of specific areas. For example, in the tourism field, selecting suitable tourist attraction attributes process is complicated as the recommendation basis for tourist attractions. In this paper, we propose the improved Attention Knowledge Graph Convolution Network model, named ($Att-KGCN$), which automatically discovers the neighboring entities of the target scenic spot semantically. The attention layer aggregates relatively similar locations and represents them with an adjacent vector. Then, according to the tourist's preferred choices, the model predicts the probability of similar spots as a recommendation system. A knowledge graph dataset of tourist attractions used based on tourism data on Socotra Island-Yemen. Through experiments, it is verified that the Attention Knowledge Graph Convolution Network has a good effect on the recommendation of tourist attractions and can make more recommendations for tourists' choices.
\end{abstract}
\section{Introduction}
With the development of the information era and the continuous improvement of web technology, people can obtain details information on scenic spots worldwide. Due to the diversity of visitors' personalized interests and the impact of information overloading. It is difficult for tourists to make a suitable choice to visit most of the neighboring tourist sites in the least amount of time. Therefore, recommendation systems have played a critical role in the tourism area by utilizing users' historical interaction information and the relationship between scenic spots\cite{b1} and making recommendations based on common preferences. Accordingly, researchers have offered a variety of academic and commercial knowledge graphs, such as NELL, DBpedia, Google Knowledge Graph, and Microsoft Satori \cite{b21}. Knowledge graphs ($KGs$) are one study interested in the side information for recommendation systems, which contain valuable facts and encode structured information about entities and their rich relations. Due to entities and their relations being highly dimensional and heterogeneous, a $KG$ is always preprocessed by knowledge graph embedding ($KGE$) methods \cite{b22}, which embeds entities and relations into low-dimensional vector spaces while preserving their inherent structure. In the traditional recommendation algorithms, the TransE algorithm\cite{b2} is employed to calculate the low-dimensional vector of the entity in the $KG$ by using the semantic relation to calculate the semantic similarity between entities and find the entity most similar to the target entity. However, the TransE algorithm can only handle simple single relationships between entities, and does not take into consideration of the multiple relations between entities. Besides, Ruihui et al.\cite{b3} introduced the combination of semantic information of items and collaborative filtering recommendation algorithm by calculating the semantic similarity between items. It also considered the relationship between users and the full utilization of similar items' characteristics; thus, they enhanced the effect of recommendations based on the $KG$.
In line with that, Jia et al.\cite{b5} proposed a network embedding method to extract the features in the knowledge graph and fully use the relationship between entities in the knowledge graph. 
On the other hand, in Xiaoyuan and Khoshgoftaar \cite{b4} the independent modeling of attribute subgraphs of different labels in the tourism knowledge graph is introduced. Thus, these models have been used to mine the semantic features of tourists, scenic spots, and other graph nodes to obtain the feature vectors of tourists and scenic spots with different characteristics. Thus, they generated the scenic spot recommendation lists by calculating the correlation between tourists and scenic spots. Moreover, Li et al. \cite{b7} introduced the scenic spot recommendation framework based on the knowledge graph, and combined the recommendation process with the knowledge graph embedding. The authors came up with a path construct for users' interest in the knowledge graph to serve as the recommendation basis.\\
Based on the analysis of actual tourism data, the recommendation algorithm based on collaborative filtering uses the interaction history of other users to recommend items to target users, ignoring the items' characteristics\cite{b8}. In contrast, the recommendation algorithm based on a knowledge graph fully takes the relationship between entities and the characteristics of entities to provide more accurate recommendations for target users. However, this recommendation algorithm only considers the first-order relationship of entities but does not consider the multi-order relationship between entities. In addition, in the professional field, especially in the tourism field, the recommendation systems effect for tourist attractions is still relatively poor. \\
 Moreover, in this research, we propose $Att-KGCN$ as a recommendation model for scenic spots according to the interaction information between tourists and scenic spots and the semantic relationship between scenic spots. Besides, constructing a knowledge graph with high-level structural information and semantic information between entities based on the relationship between scenic spots, to recommend more similar scenic spots that fall under personal interest.\\
Our main contributions in terms of designing the model architecture are as follows:
\begin{itemize}
\item Based on the $KGCN$ model, we propose an attention layer that combined among network layers to calculate the attention weight between the target scenic spot and the neighboring entities. Thus, the $Att-KGCN$ model is proposed, which can quantify the potential neighboring entities of the target scenic spots that are semantically related and dig out more tourist attractions of interest.
\item We collect data on scenic spots and tourists' historical information on Socotra Island from different websites. Thus, the knowledge graph of tourist attractions and the interaction matrix between tourists and scenic spots are constructed to recommend scenic spots.
\item The experiment results on the dataset used confirmed that the $Att-KGCN$ model performed as a recommendation system has an accurate efficacy in recommending scenic spots.
\end{itemize}
The structure of this paper is organized as follows. Section 2 introduces the research background of recommendation based on the knowledge graph and graph convolutional network. Section 3 describes the details of the proposed model employed in the study, including the phases of the model architecture. The results of the empirical experiments and discussion is highlighted in Section 4. Finally, section 5 provides the main conclusions of the study.

\section{Related Work}

\subsection{Recommendation based on the knowledge graph}
The concept of knowledge graph was proposed by Google company in 2012 \cite{b9}. Knowledge graph aims to describe various entities or concepts existing in the real world and their correlation \cite{b10}. Each entity or concept is uniquely identified and associated with each other to form a representation similar to the semantic web. Based on the attributes of the knowledge graph, many researchers have used the knowledge graph in the recommendation field and achieved good results. Traditional collaborative filtering recommendation algorithms often have problems with cold start\cite{b11} and sparse data of user-item interactions\cite{b12}. The existing recommendation methods based on knowledge graphs can be divided into the embedding and path-based methods \cite{b13}. The methods based on knowledge graph embedding\cite{b14} mainly embed entities and relationships into low-dimensional dense vectors according to the semantic relations between knowledge graphs. The represented vectors are calculated to get the similarity between entities, then make recommendations. Some knowledge graph-embedded representation models are proposed based on the Trans series for the recommendation \cite{b15}. However, they can only be used for single-step reasoning and representation, and there are still difficulties with multi-step reasoning. \\
on the other hand, the recommendation based on the knowledge graph path\cite{b15} mainly uses the knowledge related to the item in the knowledge graph. The user's path to the item is obtained through the multiple connections between the user and the item. These methods provided recommendations based on the resulting path information of users.

\subsection{Graph convolutional neural network}
The main objective of graph neural network modeling is to use graph structure to describe the information aggregation of neighboring nodes by components of the convolution operator and pooling operator \cite{b16}. The convolution operator is used to describe the local structure of nodes. In contrast, the pooling operator performs a hierarchical representation of the learning network, reducing the learning parameters and reflecting the input data's hierarchical structure. \\
By taking advantage of the fact that the graph convolutional neural network can propagate and aggregate the information between adjacent nodes in the graph data. The data of the original structure is input into the graph convolutional neural network to propagate layer by layer, and the graph structure contains the property of local parameter sharing in the convolutional neural network. With the increase of the number of propagation layers, the perception domain of each node is continuously improved, and then more information about neighbor nodes is obtained \cite{b17}. \\ 
Moreover, recommendation systems constructed by graph neural network modeling can be considered a link prediction problem of the graph signal.  They make full use of the powerful feature characterization ability of the convolutional graph network\cite{b18}. Consequently, the convolutional graph network has been proposed to solve the recommendation problem. A recommendation system is usually regarded as a complete matrix or link prediction method. Graph convolutional neural networks can well model graphs' structural attributes and node features information. Therefore, the convolutional neural network graph has played a significant role in recommendation systems and has confirmed effectiveness in various fields\cite{b20}.

\section{Methodology}
Due to the diverse attributes of tourist attractions, their choice is often full of variability, which increases the difficulty of recommending more relevant tourist attractions. Based on the recommendation of the knowledge graph and the characteristics of the graph convolutional neural network, this study proposes an improved Attention knowledge graph convolutional network to recommend tourist attractions. First, constructing a knowledge graph of tourist attractions and a visitor interaction matrix. Then, the knowledge graph convolutional network structure is created, which determines the neighboring entities of the target attractions. Thus, to perform aggregation representation, an attention mechanism is proposed to effectively identify which entities are aggregated and more similar to the target entity. Finally, the probability of tourists selecting and recommending similar attractions is calculated.

\subsection{Construction of knowledge graph of tourist attractions}
\begin{figure}
    \centering
    \includegraphics[width=8cm]{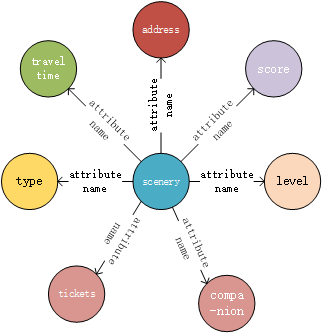}
    \caption{Travel knowledge graph}
    \label{fig:my_label1}
\end{figure}

The Google search engine and some tourist sources affiliated with the Yemeni Ministry of Tourism were employed to collect tourist attractions information on Socotra Island and information about visiting tourists. 1500 scenic spots, 2229 tourists, and 6091 score records were collected. Fig. \ref{fig:my_label1} depicts dataset details information of the scenic spots to Hadibo City in Socotra Island includes: address, score, level, tickets, travel type, travel time, travel companions, etc.

\subsection{ Attention Knowledge Graph Convolution Network Framework}
\begin{figure*}
    \centering
     \includegraphics[width=16cm]{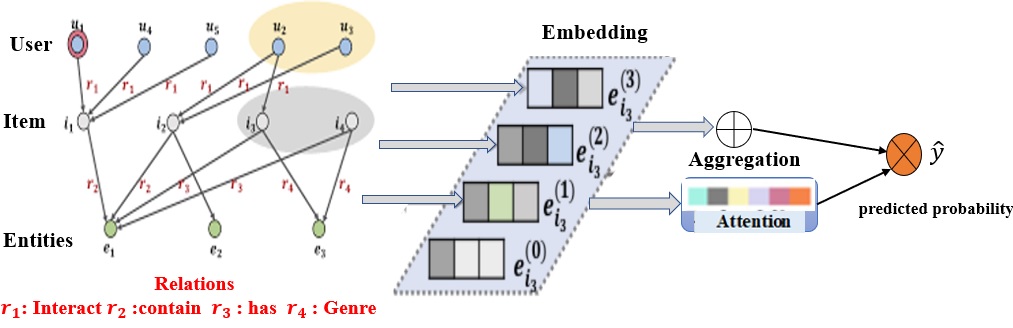}
 \caption{Framework of $Att-KGCN$ model}
    \label{fig:my_label}
\end{figure*}

The tourism knowledge graph data is used as input to the $Att-KGCN$ model. Then, the knowledge graph convolutional network is employed to represent entities as vectors by neighborhood aggregation. The low-dimensional vector embedding is performed on the acquired entities. After obtaining the aggregate representation of the target scenic spot, the attention weight between the target scenic spot and the adjacent entities is calculated. The tourists' interest scores also are computed through the relationships as the weight for the aggregation function and added to the vector representing the scenic target spot. Finally, the predicted probability of tourists' preferences for the target scenic spots is obtained. Fig. \ref{fig:my_label} shows the block diagram of the algorithm.

Further details, according to the interaction matrix between tourists and scenic spots, matrix elements are traversed to form the candidate pair ($u$,$v$) of tourist $u$ and scenic spot $v$. The scenic spot $v_i$ adjacent to the scenic spot $v$ of the first order is obtained from the knowledge graph. The $u$, $v$, and $v_i$ are represented as vectors. We use $r$ as a function to represent the relationship between scenic spot $v$ and adjacent scenic spots $v_i$ as the following:
\begin{equation}
\ r=F\left(v,v_i\right)
\end{equation}
where $v$ represents scenic spot and $v_i$ represents any adjacent scenic spot.\\
The weight of neighboring entities according to the weight of scenic spots $v$ is calculated. The greater the weight, the more likely it is to be an adjacent entity. The neighborhood entities of scenic spot $v$ are obtained.
Then, the score between the user $u$ and the relationship $r$ for scenic spots is calculated by Eq. \ref{rela}.
\begin{equation}
\ d_r=g\left(u,r\right)
\label{rela}
\end{equation}
where $d_r$ represents the importance of relation $r$ to user $u$.
Since different scenic spots have different numbers of adjacent neighborhood entities. For the ensure and efficiency of subsequent calculation, the neighborhood entities are selected in each layer with 1-order representations to obtain 2-order ones.\\
Therefore, we employ an attention mechanism to calculate the attention weight between the target entity and all adjacent entities. The neighboring first-order entities are obtained according to estimated attention weight size. The output of the attention mechanism for selected the adjoining neighborhood entities and target scenic spot entity calculated according to the formula as follows:
\begin{equation}
A_v=h^T\left(W[e_v,e_{v_i}\right])+b)
\end{equation}
Where $h$, $W$, and $b$ are all trainable parameters, $e_v$ represent the target entity, $e_{v_i}$ represent the entity adjacent to the target entity, and $A_v$ represent the attention weight of $e_v$ to $e_{v_i}$.\\
In addition, the weight of the attention mechanism between scenic spot $v$ and neighborhood entities and the relationship of tourists' scores are calculated to obtain the $K$-order neighborhood representation of scenic spot $v$ as shown in the Eq. \ref{scenic_spot}.
\begin{equation}
\begin{split}
e^u{_{(e_v,e_{v_i})}}=Softmax\left(A_v,u\right)\\
  =\exp{\left(A_v,u\right)/\sum_v{exp\left(A_v,u\right)}}
\end{split}
\label{scenic_spot}
\end{equation}
where $A_v$ represent the attention weight of $e_v$ to $e_{v_i}$ and $u$ represent user directly interacts with $v_i$.

Moreover, to aggregate the entity representation $v$ and its neighborhood representation, we used three famous functions in the knowledge graph convolution network that we will evaluate in experiments. We can describe them as follows:\\ 
(1) The summation aggregator: it performs the summation of two vectors and then a nonlinear transformation as shown in the formula as follows:
\begin{equation}
{agg}_{sum}=\sigma(w.(v+v_{s(v)}^u)+b)
\end{equation}

(2) concat aggregator \cite{b23}: it concatenates two vectors and then performs the nonlinear transformation, the formula is as follows:
\begin{equation}
{agg}_{concat}=\sigma(w . (v,v_{s(v)}^u)+b)
\end{equation}

(3) The neighbor aggregator\cite{b24}: it takes the neighborhood representation of entity $v$ as the output representation, and the formula is as follows:
\begin{equation}
{agg}_{neighbor}=\sigma(w.v_{s(v)}^u)+b)
\end{equation}
Where $s(v)$ represents the entity set adjacent to $v$, $v_{s(v)}^u$ represents the adjacent entity of the scenic spot $v$ that user $u$ directly interacts with and $w$ is the trainable parameter. \\
The final step in the proposed model aims to predict whether user $u$ has a potential interest in entity $v_i$ with which he has had no interaction.\\ Therefore, the model can learn the prediction function $ \tilde{Y}_{uv} $ that denotes the probability that user $u$ will interact with entity $v$. Besides, the final $H$ order entity of the scenic spot $v$ is expressed as $v_u$  as shown in the following formula: 
\begin{equation}
\tilde{Y}_{uv}=f(u,v_u)
\end{equation}
To investigate the effectiveness of the model, the cross-entropy was used during training as a cost function to measure the average loss between the predicted output $\tilde{y}$ and the ground truth $y$ as the following formula:
\begin{equation} 
\mathcal{L}= \sum^n_{i=1}{- \ln \sigma (y{(u,v_i)}-\tilde{y}{(u,v'_i)}) } 
\end{equation}
Where $\sigma(\cdot)$ is the sigmoid function.

\section{Experimental Setup}
In this section, we analyze numerous experiments we conducted with the Att-KGCN model on a real-world dataset. We evaluate the proposed model by tuning hyper-parameters and selecting the optimal values.
\subsection{The dataset}
The data on scenic spots and the rating data of tourists for scenic spots in Socotra Island are collected as experimental data.
The data contain several attributes such as location, interest degree, type, ticket price of scenic spots, etc. The score of interest of tourists is recorded as an attribute of the interaction between tourists and scenic spots. The data contains missing values and interfering data caused by the anomaly in some columns or rows. Therefore, when we analyze and filter directly without considering these exceptions and errors, it will directly or indirectly affect the accuracy of the results. Accordingly, data preprocessing steps are required to ensure the accuracy of the results.  The first phase is to exclude the records containing no data. The next stage is to delete columns with data noise, cause the model's performance inaccuracy, and remove all the empty columns. In addition, in the knowledge graph, to represent the relationship between two scenic spots they must have the same attribute values. Therefore, more than 25,000 triples are obtained by constructing a tourism knowledge graph. The basic statistics of the dataset used are shown in Table \ref{tbl1}.
During the experiment, the data set was divided into the training set, verification set, and test set according to 7:1:2, and the training was carried out with a batch size of 32. 
\begin{table}[ht]
\caption{Description of Dataset.}
\label{tbl1}
\begin{center}
\begin{tabular} {c|c}
\hline
\multicolumn{2}{ c }{\textbf{Attributes of dataset}}\\\hline
users &	  51 \\ \hline 
items  &   5210 \\ \hline
entities  &	  9982  \\ \hline
relations	  & 14  \\ \hline
$KG$ triples  &	25575 \\ \hline
\end{tabular}
\end{center}
\end{table}
\subsection{Evaluation Metrics}
In the recommendation system, evaluation metrics are usually used to measure the efficiency of the proposed model. The commonly used evaluation metrics include area under the $ROC$ curve ($AUC$) and F$-$score ($F1$).\\

1) $AUC$ reflects the ranking quality predicted by the model, that is, the proportion of positive examples in front of negative examples. 
The larger the proportion is, the better the prediction effect of the model is. It takes the numbers between [0,1].

2) F$-$score($F1$) is the weighted average of accuracy and recall rate, which uniformly reflects the effect of model recommendation.
\begin{equation}
F1 = 2 \times \frac{precision \times recall}{precision + recall}
\end{equation}

3) Accuracy is used to measure how good the model is for the correct prediction of two groups, for users (non-engaged or engaged).

\begin{table}[ht]
\caption{Hyper-parameters used in the $Att-KGCN$ Model}\label{tbl2}
\begin{center}
\begin{tabular}{c|c}
\hline
\multicolumn{2}{c}{\textbf{Hyper-parameters of $Att-KGCN$  Model}}\\\hline
$K$ &	 14\\ \hline 
$d$  &  32\\ \hline
$H$  &	2\\ \hline
$\lambda{}$	 & 2e-2\\ \hline
$\eta$ & 1e-2\\ \hline
batch-size  &32\\ \hline
\end{tabular}
\end{center}
\end{table}
\subsection{Parameter Settings and Experimental Results}
We implement our $Att-KGCN$ model in TensorFlow with hyper-parameters determined by optimizing $AUC$ and F1-Score metrics on a validation set as shown in Table \ref{tbl2}. The $K$ represents neighbor sampling size, which refers to the count of neighbors that are aggregated by the aggregation function. Since most nodes have a huge number of neighbors, aggregating all the neighbors is neither possible nor effective since it leads to the problem of overfitting. So the K value needs to be selected such that it both captures important neighborhood information as well as prevents a noise problem. In our experiment, the number of $K$ is chosen as 4, 8, 16, 20, 24, and 28 as shown in Table \ref{tbl3}.  As seen from the results of the experiments on a validation dataset,  the optimal value when k was set to 8 neighbor entities, then the next when was 16. Because when $K$ is too small, there is not enough capacity to accommodate the adjacency information. Conversely, when $K$ is too large, it is easy to contain a lot of redundant information thus reducing the performance of the model.

The $d$ is the dimension of the embeddings, which represents the vector dimension for a node, which can be obtained by performing convolution operations on graphs like GraphConv to generate the representation for the node without having to feature engineering techniques. To assess the effect of the embedding dimension $d$, the embedding dimension $d$ was assigned as 8, 16, 32, 64, and 128, whereas the other parameters were fixed as depicted in Table \ref{tbl4}. 
As we note in the table results, The best value of $AUC$ was when the embedding dimension was 16 but the best value of $F1-score$ when the d is 8. 

The $H$ is the depth of the receptive field which means the ability of the model to capture long-term relations in the KG. When the $H$ value is set to 1, the model captures directly connected entities. The higher the value of $H$, the more distant interests of the user can be captured. Therefore, we must select this value carefully because values too large may capture even relations that are not interesting to the user.  In our experiment, the $H$ was chosen as 1, 2, 3, 4, and 6 as reported in Table \ref{tbl5}.  The experimental results demonstrate that when the depth of the acceptance domain is 1, the effect is the best. Besides, when the depth of the acceptance domain was selected 2 the resulting values indicated a positive effect on the model's performance.  On the other hand, when the value of H was more than 2, the results showed a deterioration in the performance of the model. This can be justified because the larger the representation dimension, the more user and entity information will be represented. It leads to slow convergence, and the model tends to overfit. In addition, If the relationship chain is too long, it will bring great interference to the model, and there will be no basis for inferring the similarity between items.
\begin{table}[htbp]
\centering
\caption{$AUC$ and $F1$-$score$ Results on different $K$ for $Att-KGCN$ model.
    The best performance is highlighted in bold.}\label{tbl3}
    \begin{tabular}{lccc}
        \hline
        \textbf{$K$} & \textbf{$AUC$}& \textbf{$F1$-score} \\ \hline
              2 & 0.924  &  0.895  \\\hline   
              4	& 0.9467	&0.9045 \\\hline 
              8	&  \bfseries 0.9778	& \bfseries 0.9523  \\\hline
             16 &	0.9769 &	0.9285 \\\hline
             20 &	0.976 &	0.948  \\\hline
             24 &	0.964 &	0.938  \\\hline
            28  &	0.959 &	0.932 \\\hline
    \end{tabular}
\end{table}

\begin{table}[htbp]
\centering
\caption{$AUC$ and $F1$-$score$ Results on different d for Att-KGCN model.
 The best performance is highlighted in bold.} \label{tbl4}
    \begin{tabular}{lccc}
        \hline
        \textbf{d} & \textbf{AUC}& \textbf{F1-score} \\\hline
              8	& 0.973	& \bfseries 0.944  \\\hline   
              16 &	\bfseries 0.976 &	0.923 \\\hline 
             32 &	0.941 &	0.897  \\\hline
             64	& 0.936	& 0.892  \\\hline
            128	& 0.884	& 0.831  \\\hline
    \end{tabular}
\end{table}
\begin{table}[htbp]
\centering
\caption{$AUC$ and $F1$-$score$ Results on different H for Att-KGCN model.
    The best performance is highlighted in bold.}\label{tbl5}
    \begin{tabular}{lccc}
        \hline
        \textbf{H} & \textbf{AUC}& \textbf{F1-score} \\\hline
              1	& \bfseries 0.978	& \bfseries 0.952  \\\hline  
              2	& 0.965	& 0.913 \\\hline
            3 &	0.884 &	0.849 \\\hline
            4 &	0.809 &	0.795 \\\hline
            6 &	0.664 &	0.60 \\\hline
    \end{tabular}
\end{table}
\begin{table}[htbp]
\centering
\caption{$AUC$ and $F1$-$score$ Results on different aggregation functions for Att-KGCN model.
    The best performance is highlighted in bold.}\label{tbl6}
    \begin{tabular}{lccc}
        \hline
        \textbf{$Function$} & \textbf{AUC}& \textbf{F1-score} \\\hline
              ${agg}_{sum}$	& \bfseries 0.976	& \bfseries 0.953  \\\hline  
              ${agg}_{concat}$	& 0.961	& 0.92 \\\hline
            ${agg}_{neighbor}$ &	0.880 &	0.851 \\\hline
    \end{tabular}
\end{table}

\begin{figure}[htbp]
\centerline{\includegraphics[scale=0.5]{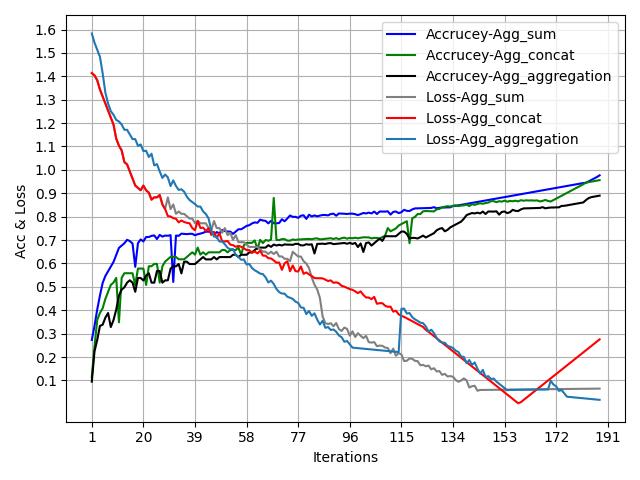}}
\caption{$Accuracy$ and $Loss$ Curve of the $Att-KGCN$ Model through training.}
\label{Accloss}
\end{figure}

Given the experiments result, the optimal values for $AUC$ on the evaluation dataset were $K$ = 8, $d$ = 16, and $H$ = 1,2. The optimal values for the F1-score results are $K$ = 8, $d$ = 8, and $H$ = 1,2. 
In line with that, we conduct experiments on the most suitable values (i.e., $k$=8, $d$=16, $H$=2) with the three aforementioned aggregation functions. Other experimental parameters are tuned such as $\lambda{}$ is  $L2$ regularizer weight, $\eta$ is the learning rate, and the batch size fixed is 32. The $H$ is set to 2 to make the model captures 2-order representations for connected entities. All trainable parameters are optimized by the Adam algorithm. The 20 iterations are carried out, and the changes for $AUC$ and $F1$ values and the average performance are recorded as shown in Table\ref{tbl6} and Fig.\ref{Accloss}. As we see from Fig.\ref{Accloss}, which reports comparisons of the visualization analysis on the training, validation, and testing dataset in accuracy and loss metrics. In the early training steps, the accuracy and loss of the model obtained have not fit well enough with three different aggregation functions. With progress in implementation time, the $Att-KGCN$ model with ${agg}_{sum}$ alleviates the fluctuations and the overfitting issue throughout the training stage and achieves better results.

\begin{figure}[htbp]
\centerline{\includegraphics[scale=0.5]{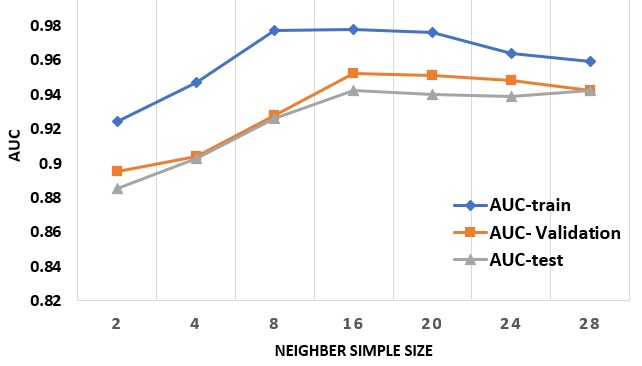}}
\caption{$AUC$ Curves of the $Att-KGCN$ Model with different neighbor sampling sizes on training, testing, and validation dataset.}
\label{AUC}
\end{figure}

Moreover, when the aggregation function is employed in the proposed model, not only the tourist relationship score is used as the weight for aggregation, but also the attention weight between the target entity and neighboring entities is also calculated. As well as it also considers full personalized preferences, improving the recommendation's effects.  As we see from the table, the proposed model with ${Agg}-{sum}$ performs best. This indicates that capturing users’ personalized preferences and semantic information about the KG does benefit the recommendation. In contrast, the proposed model with ${agg}_{neighbor}$ was worse. This may be because the neighbor aggregator only employs the neighborhood representation, thus losing helpful information from the entity. Besides, We can see Fig.\ref{AUC} shows $AUC$ results of $Att-KGCN$ with different neighbor sampling sizes on the training, testing, and validation dataset. The $K$=8 has the highest value. On the contrary, $K$=2 has the smallest value.

\begin{table}[htbp]
\caption{Compression Performance of $Att-KGCN$ with $KGCN$ }\label{tbl7}
\begin{center}
\begin{tabular}{lccc}
\hline
model & $AUC$ & $F1-score$\\ \hline 
$KGCN$ &0.96 &	0.92 \\ \hline 
$Att-KGCN$ &0.98 &	0.95\\ \hline
\end{tabular}
\end{center}\label{comp}
\end{table}
To be more accurate in efficiency and effectiveness, we conducted the experiments on KGCN without an attention mechanism. It's observed that the proposed model performs better than the one that doesn't include an attention mechanism as depicted in Table \ref{comp} for $AUC$ and $F1-score$ metrics values.
\section{Conclusion}
he paper introduced $Att-KGCN$  as a recommended system for scenic spots, and the higher-order structure information and semantic information in the knowledge graph of scenic spots are fully considered to predict the probability of tourists choosing target scenic spots. Due to various neighborhood entities in varying degrees for a given target site. The attention mechanism is proposed to adjust target spots' neighborhood weights and the target entity and its neighborhood spot target are obtained by the aggregation functions. Finally, the probability of tourists choosing the scenic spot is calculated. This research uses a knowledge graph dataset of tourist attractions based on tourism data on Socotra Island-Yemen. The experimental results proved $KGCN-Att$ model led to digging out more scenic spots that tourists are interested in and improved scenic spots recommendation performance. 

However, the weakness is that the relationships between scenic spots are still relatively small, so more relationships must be considered to further improve the recommendations' interpretability.

\bibliographystyle{IEEEtran}
\bibliography{Refrence}

\end{document}